# Neutron diffraction and symmetry analysis of the martensitic transformation in Co-doped Ni$_2$MnGa

Fabio Orlandi,[1, ∗] Aslı Çakır,[2] Pascal Manuel,[1] Dmitry
D. Khalyavin,[1] Mehmet Acet,[3] and Lara Righi[4, 5]

[1]*ISIS Facility, Rutherford Appleton Laboratory - STFC,
OX11 0QX, Chilton, Didcot United Kingdom*

[2]*Muğla Sıtkı Koçman University, Department of Metallurgical
and Materials Engineering, 48000 Muğla, Turkey*

[3]*Faculty of Physics and Center for Nanointegration (CENIDE),
Universität Duisburg-Essen, D-47048 Duisburg, Germany*

[4]*Department of Chemistry, Life Sciences and Environmental Sustainability,
University of Parma, Parco Area delle Scienze 17/A, 43124 Parma, Italy*

[5]*IMEM-CNR, Parco Area delle Scienze 37/A, 43124 Parma, Italy*

(Dated: March 17, 2020)

# Abstract

Martensitic transformations are strain driven displacive transitions governing the mechanical and physical properties in intermetallic materials. This is the case in $Ni_2MnGa$, where the martensite transition is at the heart of the striking magnetic shape memory and magneto-caloric properties. Interestingly, the martensitic transformation is preceded by a pre-martensite phase, and the role of this precursor and its influence on the martensitic transition and properties is still a matter of debate. In this work, we report on the influence of Co doping ($Ni_{50-x}Co_xMn_{25}Ga_{25}$ with x = 3 and 5) on the martensitic transformation path in stoichiometric $Ni_2MnGa$ by neutron diffraction. The use of the superspace formalism to describe the crystal structure of the modulated martensitic phases, joined with a group theoretical analysis allows unfolding the different distortions featuring the structural transitions. Finally, a general Landau thermodynamic potential of the martensitic transformation, based on the symmetry analysis is outlined. The combined use of phenomenological and crystallographic studies highlights the close relationship between the lattice distortions at the core of the $Ni_2MnGa$ physical properties and, more in general, on the properties of the martensitic transformations in the Ni-Mn based Heusler systems.

## I. INTRODUCTION

Martensitic transformations, diffusion-less, displacive and first order transitions from a high symmetry austenite phase to a low symmetry martensite phase,[1] are strain-driven phase transitions observed in different classes of materials.[1, 2] The transformation is characterized by the appearance of strong lattice strains frequently joined with shuffles and/or shears of specific lattice planes resulting in macroscopic changes of the crystal shape and of the material microstructure.[1] Diverse and striking physical phenomena occur at the martensitic transformation ranging from enhancement of the material's mechanical strength,[1, 2] shape memory effects,[3] magneto- and mechano-caloric effects[4, 5] and even exotic new topological states.[6] The knowledge of martensite symmetry and how the distor-

---

* corresponding author:fabio.orlandi@stfc.ac.uk

tions develop during the transformation are pivotal in understanding the functional physical properties characterizing such materials.[7, 8]

Heusler alloys are a wide class of materials showing martensitic transformations induced by different external stimuli such as temperature-changes,[9–13] magnetic-field [14, 15] as well as applied strain and pressure.[13, 16–18] Moreover, the Heusler structure is able to host a large variety of chemical species by adjusting the lattice symmetry and distortions,[13, 19–24] resembling the structural flexibility observed in perovskite oxides. The ideal cubic structure of austenite can undergo symmetry-breaking transitions showing close analogies to perovskite distortions.[19] Let us consider a martensitic transformation that takes place from a high-temperature cubic L$2_1$ austenite phase (space group $Fm\bar{3}m1'$) to a low-temperature tetragonal non-modulated L$1_0$ martensite phase ($I4/mmm1'$). The martensitic distortion is usually driven by a tetragonal strain acting on the Mn coordination polyhedra like a Jahn-Teller distortion as shown in Fig. 1. This local effect is known to occur in perovskite oxides, and was also observed in the Ni$_2$MnGa system from neutron diffraction experiments and asserted as a band Jahn-Teller effect.[25, 26] For some compositions, the martensitic transformation is also associated with a periodic shift of specific lattices planes that can give rise to commensurate and incommensurate structure modulations.[11, 21–23, 27] These displacive modulations can be viewed as a way for the lattice to maintain the proper geometrical bond requirements and can be compared with the characteristic oxygen tilting patterns in perovskite materials. [28, 29] Many efforts of the research community were dedicated to the understanding of the various lattice distortions in perovskite materials[30] and how the interplay between them leads to functional properties like multiferroicity.[31, 32]

In this paper, we will apply the formalisms generally used in the perovskite materials to the martensitic transformation in Ni$_2$MnGa Heusler alloys. Depending on composition, Ni$_{50}$Mn$_{45-x}$Z$_x$ (Z: Ga, In, Sb, Sn) alloys show a wide range of multifunctional properties related to the matensitic transformation, ranging from magnetic shape memory effects,[14, 33, 34] giant magneto- and barocaloric effects [4, 5, 35, 36] to exotic magnetic properties.[12, 37] These technologically relevant properties arise from the relationship between different degrees of freedom in the material: magnetism, spontaneous strains, displacive modulation and the interplay of these with the system entropy.[5] All resemble the interaction of the different ferroic orderings in multiferroic materials, that is strictly linked to the coupling of the different nuclear and magnetic distortions.[38] It is therefore interesting to investigate the lattice and magnetic distortions and their coupling. To fulfill this purpose, we start from the experimental observation of the change in the martensitic transformation path in stoichiometric $Ni_2MnGa$ induced by small Co doping. $Ni_2MnGa$ can be considered the archetype of Ni-Mn based Heusler alloys, and has been widely studied in literature. It undergoes a martensitic transformation from a cubic austenite phase, described with the $Fm\bar{3}m1'$ space group, to a 5M modulated martensite structure below 210 K.[22] The martensitic transformation is preceded by a "premartensite" phase below 260 K.[10, 26, 39] This phase can be described as a modulated austenite phase with an incommensurate modulation vector close to the commensurate ($1/3$ $1/3$ 0) value (defined with respect to the cubic parent structure). The nature of this premartensite phase and its role in the martensitic transformation are still a matter of debate,[40–42] justifying the need of a detailed symmetry analysis of the transformation.

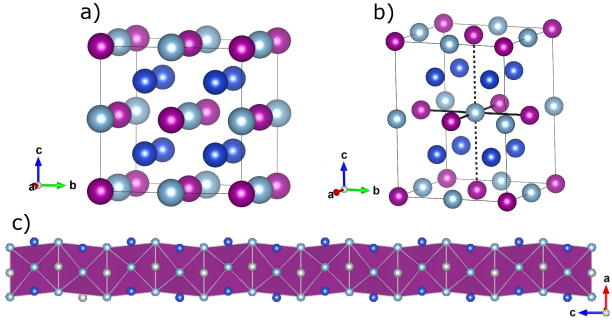

FIG. 1. a) Crystal structure of the austenite phase in the Heusler alloys $X_2YZ$ (X blue, Y purple, Z gray). (b) Example of the tetragonal strain effect on the coordination environment the Z (or X) sites that resemble the Jahn-Teller effect in perovskites. (c) Example of shuffle that resemble the octahedral tilting in perovskite oxides.

Two main theories are usually considered when trying to explain the martensitic transformation in $Ni_2MnGa$ and related compounds: a soft phonon model[40] and an adaptive one.[41, 43, 44] In the latter, the modulated martensite phase is described as a long-range ordered arrangement of nano-domains of the $L1_0$ tetragonal structure. The strict geometrical relation between the austenite and martensite lattices, due the conservation of the habit plane [44], and the minimization of the elastic energy [43] restrict the periodicity of the modulated phases. The adaptive model considers the modulated martensite phase as a metastable phase and the tetragonal non-modulated $L1_0$ phase as the only thermodynamically stable martensitic phase.[43, 44]. This model is able to correctly describe the mechanical properties of the martensite phases together with the magnetic anisotropy. [41, 45, 46] Nevertheless it also shows some limitation: (i) the observation of incommensurate structures is not well justified and are explained by the presence of stacking faults in the ordered twinning pattern; (ii) it does not fully account for the presence of the pre-martensite phase, in which the tetragonal strain is not quantifiable but the modulation satellites are still present; (iii) it does not explain why the nano-twinning need to be periodic on a large scale, which is needed to explain the observation of sharp Bragg reflections.

The soft-phonon model take into account the experimentally observed softening in the $[\zeta\zeta 0]$ TA$_2$ phonon branch at $\zeta \approx 0.33$. [40] This theory has been mainly developed in the ferroelectic community [47] and it is based on the observation that one, or a limited number of phonons, posses a lower frequency that decreases with temperature, eventually becoming unstable at the transition. In its first description this theory required a complete softening of the phonon mode and a classical second order phase transition. [47–49] This is not the case in Ni$_2$MnGa where an incomplete softening of the unstable phonon and a first order transition are observed. Nevertheless, Krumhansl and Gooding[49] show that the requirement of a complete softening of the phonon mode and a second order transition is not a strict requirement and can be overcome, for example, considering a coupling between the soft phonon mode and spontaneous strains.[50] In the case of Ni$_2$MnGa this theory is criticized on the lack of a complete phonon softening and on the fact that it fails to explain the presence of the precursor phase.[40] Recently Gruner *et al.* [44] proposed a stricter relation between the phonon softening and the adaptive nano-twinning.

In this work we will show an experimental proof of a direct coupling between the incommensurate modulation (the soft-phonon mode) and the tetragonal strain. We will also provide a Landau thermodynamical potential based on the symmetry of the parent austenite phase and the daughter martensite phases that formally shows the phenomenological effect of the latter coupling on the macroscopic properties of this systems, strengthening the soft-phonon model. These conclusions are based on neutron diffraction data collected on Co doped Ni$_2$MnGa alloys. The macroscopic properties and the symmetry of the martensite phase in Ni$_2$MnGa are very sensitive to small variations of the composition,[11, 12, 21, 23, 24, 36, 51, 52] so that varying the concentration gives the opportunity to gain insights on the martensitic transformation. In particular, the substitu-tion of Co in the Ni sublattice influences both magnetic and structural properties of the alloy,[12, 24, 36, 51, 52] allowing the direct observation of the coupling between the different degrees of freedom.

## II. EXPERIMENTAL METHODS

$Ni_{50-x}Co_xMn_{25}Ga_{25}$ samples ($x = 3, 5$ nominal) were prepared by arc melting of high purity elements (99.99%). Ingots were sealed in quartz tubes under 300 mbar of Ar and then annealed at 1073 K for 5 days. The sample-compositions and uniformity were determined by energy dispersive x-ray (EDX) analysis using a scanning electron microscope. The stoichiometries are $Ni_{46.04}Co_{2.59}Mn_{25.96}Ga_{25.42}$ and $Ni_{44.54}Co_{4.97}Mn_{25.35}Ga_{25.13}$. The uniformity is within 0.05%.

Temperature-dependent magnetization, M(T), measurements were performed with a superconducting quantum interference device magnetometer (SQUID) under a 5 mT applied-field in the temperature range $5 \leq T \leq 650$ K with a 4 Kmin$^{-1}$ sweeping rate. An oven was attached to the magnetometer for the high-temperature measurements. The measurements were carried out in a zero-field-cooled (ZFC), field-cooling (FC) and field-cooled-warming sequence (FCW).

The neutron powder-diffraction data were collected on the time-of-flight WISH diffractometer at the ISIS facility (UK).[53] The powder-sample was loaded in a thin vanadium sample-holder and measured between 5 and 500 K with a hot-stage closed-circuit refrigerator. Rietveld refinements were performed using the Jana2006 software,[54] and group theoretical calculations were carried out with the help of the ISOTROPY Suite.[55, 56] The symmetry-mode analysis [30, 57] of the martensitic sequence of transitions in $Ni_{2-x}Co_xMnGa$ is performed with the help of the ISODISTORT software[55] using the refined paramagnetic (PM) austenite phases and the relative distorted structures. The Landau thermodynamic potential was constructed with the help of the ISOTROPY suite[56] and INVARIANT software's.[58]

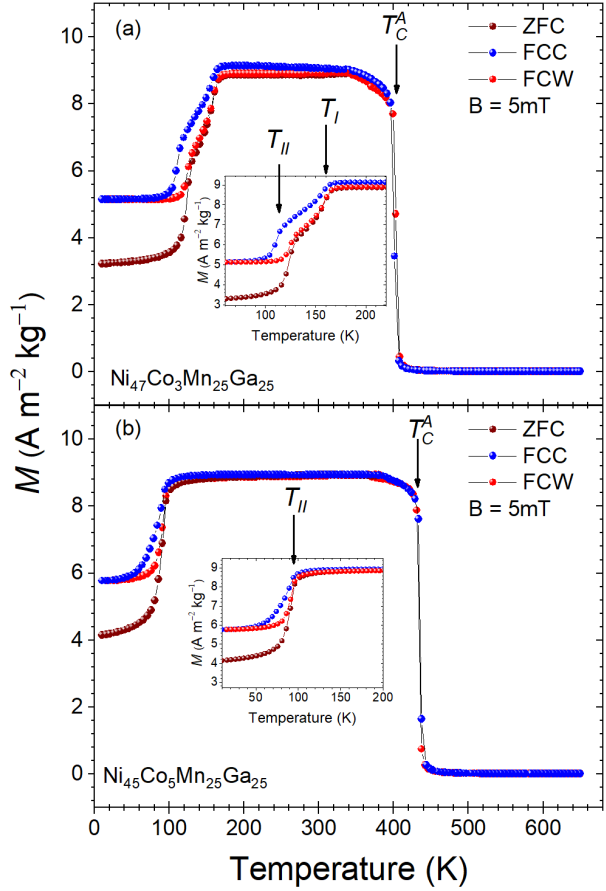

FIG. 2. Magnetization measurements of (a) $Ni_{47}Co_3Mn_{25}Ga_{25}$ and (b) $Ni_{45}Co_5Mn_{25}Ga_{25}$ measured in an applied field of 5mT following a Zero field cooled (ZFC), Field Cooled (FC), and Field Cooled Warming (FCW) procedure.

## III. RESULTS

### A. Magnetization

M(T) obtained in 5 mT applied field for the two samples studied in the present work are shown in fig. 2. All compositions show a sharp transition above 400 K indicating the ferromagnetic (FM) Curie temperature of the austenite phase ($T_C^A$). As already reported by Kanomata et al.,[52] $T_C^A$ of the austenite phase increases with increasing Co concentration. The estimated transition temperatures, 405 and 437 K for x = 3 and 5, respectively, are in agreement with those given in reference [52]. Both compounds undergo at low temperatures a martensitic phase transition, which is accompanied by a hysteresis between the FC and FCW M(T) data. The ZFC state is obtained by cooling the sample from 380 K to the lowest temperature. Since 380 K corresponds to a temperature within the FM state for these two samples, and not to a temperature in the PM state, M(T) does not become zero at the lowest temperature, and a certain finite value remains. Nevertheless, the splitting between the ZFC and FCW curves indicates the presence of mixed magnetic exchange with competing non-FM and FM components. The sample with x = 3 shows two features in M(T) around $T_I \approx$ 160 K and $T_{II} \approx$ 120 K. The sequence of the transitions resembles the behavior observed for the $Ni_2MnGa$ composition which is characterized by a pre-martensitic transition around $T_I \approx$ 260 K followed by the martensitic transformation around $T_{II} \approx$ 210 K. [39] On the other hand the sample with x = 5 exhibits a single transition around 90 K. The details of these transitions are examined more closely with neutron diffraction experiments presented below.

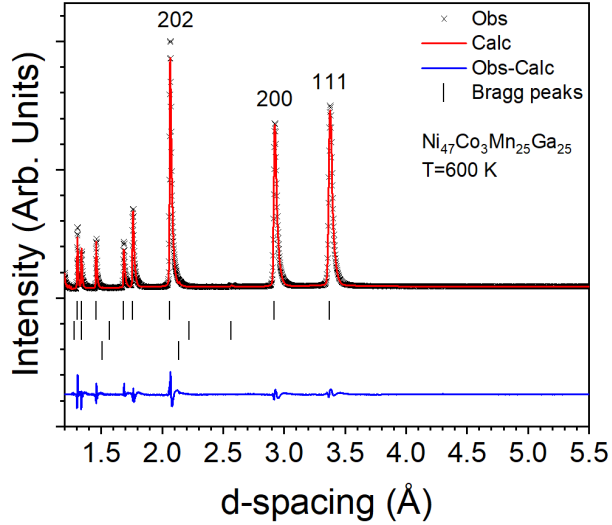

FIG. 3. Rietveld plot of the austenitic phase at 600 K in the $Fm\bar{3}m1'$ space group for the $Ni_{47}Co_3Mn_{25}Ga_{25}$ compound. Observed (x, black), calculated (red line), and difference (blue line) patterns are shown. The tick marks indicate the Bragg reflection positions. The agreement factors are $R_p$=3.8% and $R_{wp}$=3.9%. The second and third tick-marks-rows indicate the reflections from MnO and V respectively.

## B. Neutron diffraction

### 1. Paramagnetic and ferromagnetic austenite phases

In the paramagnetic state, above $T_C^A$, all the samples are in the cubic $L2_1$ phase. Diffraction data obtained at 600 K show very similar patterns for all samples. We show as example the data obtained for $Ni_{47}Co_3Mn_{25}Ga_{25}$ in fig. 3. We obtain good reliability factors of the Rietveld refinements by adopting the $Fm\bar{3}m1'$ symmetry. The chemical compositions obtained from structural refinements (by taking advantage of the good contrast of the scattering cross-section between the elements, with $b_{Mn}$ = -3.73 fm, $b_{Ni}$ = 10.3 fm $b_{Co}$ = 2.73 fm, and $b_{Ga}$ = 7.28 fm) are in agreement with the EDX measurements for all investigated compounds (see table S I to S V in the supplementary materials). [59]

In fig. 4, we show the neutron-diffraction data for $Ni_{47}Co_3Mn_{25}Ga_{25}$ and $Ni_{47}Co_5Mn_{25}Ga_{25}$ obtained at 300 K in the ferromagnetic phase. We observe below $T_C^A$ a strong temperature-dependence of the cubic 111 reflection intensity and the absence of extra reflections indicating the presence of a k = (0 0 0) propagation vector for the FM austenite phase (insets fig. 4(a) and 4(b)). The $T_C^A$ obtained as 431(3) K and 449(7) K are slightly higher than the values obtained from M(T). The magnetic symmetry analysis in the FM state was performed by referencing to the PM austenite structure and the observed propagation vector; assuming magnetic ordering of both Ni and Mn sites. The details of the analysis are given in Appendix A.

It is worth stressing here that although the resolution of the present diffraction experiment is not sensitive enough to detect a possible magneto-elastic coupling, the occurrence of very small lattice distortions cannot be excluded a priori. The correct symmetry of the magnetic state is assigned by combining the knowledge of the magnetic anisotropy and symmetry information associated with the phase transition. Firstly, from the integrated intensity of the 111 reflection as a function of temperature and from the absence of thermal hysteresis around $T_C^A$ in M(T) (fig. 2), it is evident that the magnetic transition progresses as second order. Secondly, literature related to prototypical $Ni_2MnGa$ and off-stoichiometric compositions indicate the occurrence of a uniaxial type of ferromagnetism along one of the fundamental $[001]_C$ crystallographic axis.[60] These evidences are consistent only with the $I4/mm'm'$ magnetic space group. For all compositions, the refinements of the tetragonal magnetic

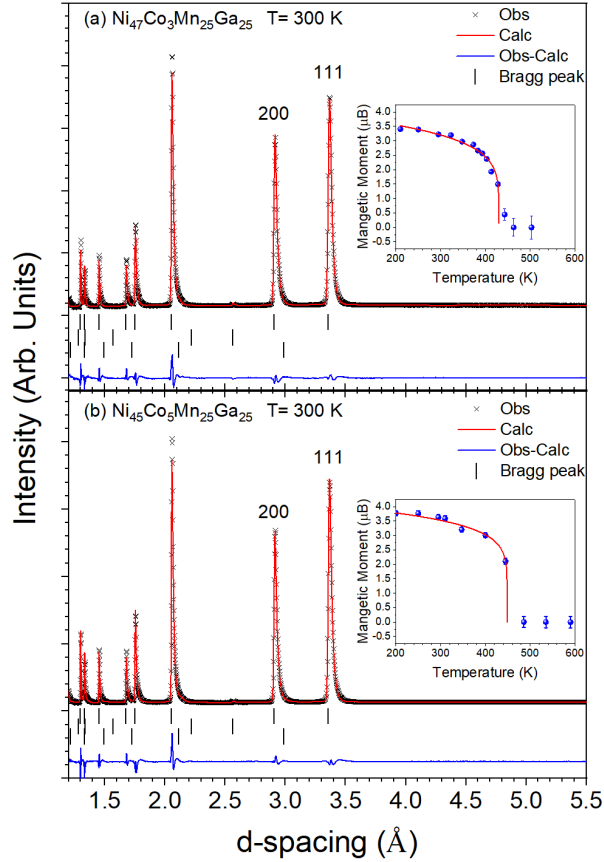

FIG. 4. Rietveld plot of the ferromagnetic austenitic phase at 295 K in the $I4/mm'm'$ magnetic space group for the (a) $Ni_{47}Co_3Mn_{25}Ga_{25}$ and (b) $Ni_{45}Co_5Mn_{25}Ga_{25}$ composition. Observed (x, black), calculated (red line), and difference (blue line) patterns are shown. The tick marks indicate the Bragg reflection positions of the main phase (first row), MnO and V (second and third row respectively). The agreement factors are $R_p$ =4.8 % and $R_{wp}$=3.9 % and $R_p$=3.7 % and $R_{wp}$=3.8 %. The inset shows the evolution of the refined magnetic moment as function of temperature pointing out the continuous character of the transition, the red line shows the best fit of the data with a power law returning $T_C^A$ of 431(3) K with critical exponent $\beta = 0.24(3)$ for $Ni_{47}Co_3Mn_{25}Ga_{25}$ and $T_C^A$ of 449(7) K and $\beta = 0.25(4)$ for $Ni_{45}Co_5Mn_{25}Ga_{25}$

space group indicate the presence of ordered magnetic moment only on the Mn sublattices along the $[001]_C$ direction. The moment on the Ni/Co sites can be considered as negligible or below the sensitivity of our diffraction experiment.

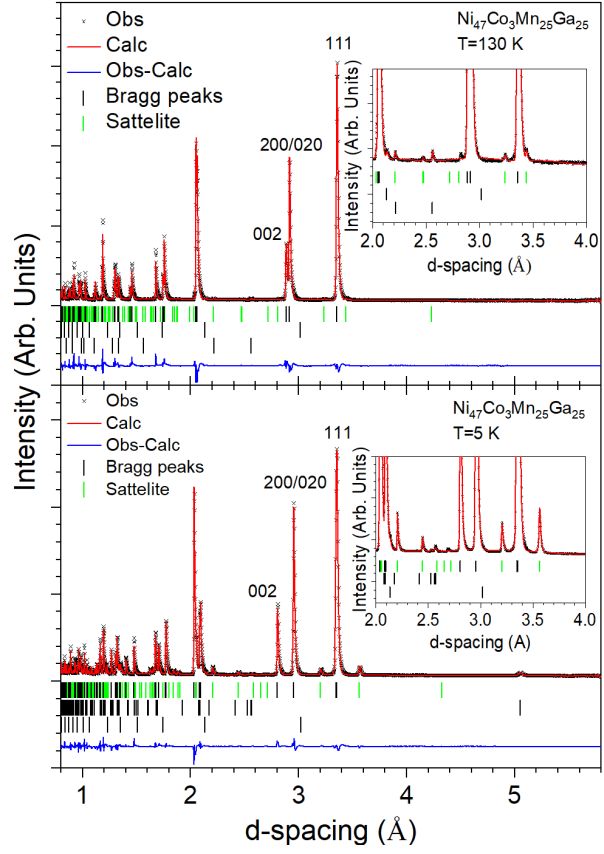

FIG. 5. Top) Rietveld plot of the first martensitic phase at 130 K in the $Im'mm'(00\gamma)s00$ magnetic superspace space group for the Ni$_{47}$Co$_3$Mn$_{25}$Ga$_{25}$ composition. The agreement factors are R$_p$=5.4 % and R$_{wp}$=6.8 %. The inset shows a zoom of the diffraction pattern highlighting some satellite reflections. Bottom) Rietveld plot of the second martensitic phase at 5 K in the $Im'mm'(00\gamma)s00$ magnetic superspace space group for the Ni$_{47}$Co$_3$Mn$_{25}$Ga$_{25}$ composition, The agreement factors are R$_p$=4.04 % and R$_{wp}$=5.05 %. The inset shows a zoom of the diffraction pattern highlighting some satellite reflections. In both panels observed (x, black), calculated (line, red), and difference (line, blue) patterns are reported. The first tick marks row indicates the Bragg reflections position of the martensite phase with the main reflections in black and the first order satellite reflections in green. The second and third tick marks rows indicate the reflections from the MnO magnetic impurity and V respectively.

2. *Temperature-induced martensitic transformation*

Following the features related to the martensitic transitions in M(T) in fig. 2, neutron diffraction measurements were performed to study in more detail the transitions.

*a.* $Ni_{47}Co_3Mn_{25}Ga_{25}$   Below T$_I$ it is possible to observe a splitting of the 200 cubic reflection characteristic of a tetragonal distortion (see fig. 5) as confirmed by whole-pattern Le Bail refinement. Besides the lattice distortion, the rising of weak reflections indicate that the new phase is modulated (see inset fig. 5). This whole set of new peaks can be indexed based on a pseudo-austenitic lattice with a modulation vector $q \approx$ (0.3 0.3 0). The values of the propagation vector and the weak intenities of the satellite reflections are characteristic of the pre-martensitic transformation as observed for Ni$_2$MnGa [10, 25, 39, 61], NiTi [62] and Ni$_{1-x}$Al$_x$ alloys.[63] However, contrary to the 3% Co sample, the Ni$_2$MnGa system, as well as NiTi and Ni$_{1-x}$Al$_x$, does not show any strain in the pre-martensite phase even with high-resolution synchrotron data.[10, 39]

The modulated state for $T_{II} < T < T_I$ is described within the superspace formalism.[64–66] The observed systematic absences of the main and satellite reflections allows to undoubtedly identify the $Immm(00\gamma)s00$ superspace group as the correct symmetry to describe the crystallographic structure. No evident changes in the magnetic contribution to the diffraction pattern were observed across the T$_I$ transition, indicating small changes in the ferromagnetic structure. Previous works indicate, in fact, that the b-axis of the martensite cell described in this work (corresponding to the cubic [001] direction) is the magnetic easy axis.[60, 67] A ferromagnetic ordering of the Mn spins along the b direction corresponds to the magnetic superspace group $Im'mm'(00\gamma)s00$. The refinement with the aforementioned symmetry at 130 K is shown in fig. 5(a) whereas crystal data and atomic positions are reported in table S VI and S VII respectively.[59]

The refined modulated structure is characterized by a sinusoidal modulation of the x-position of all the atoms present in the structure with amplitudes very similar to the one observed in the pre-martensite phase of the stoichiometric Co-free compound[10] (see table S VII[59] and fig. 6). Although the lattice symmetry is orthorhombic, the refinement of the cell parameters converged to a tetragonal metric with equal amplitude of the incommensurate displacement for the Mn and Ga site. As for the austenite phase the ordered moment on the Ni site can be considered zero within $3\sigma$. It is worth stressing that neutron diffraction allows to determine only the average values of the ordered magnetic moment on the crystallographic site and not the single value for Ni and Co. Kanomata *et al.*[52] suggest, from electronic-structure calculation, the value of $\approx$ 1 $\mu_B$ per Co atoms and $\approx$ 0.25 $\mu_B$ per Ni that leads to an average moment on the site of $\approx$ 0.3 $\mu_B$.

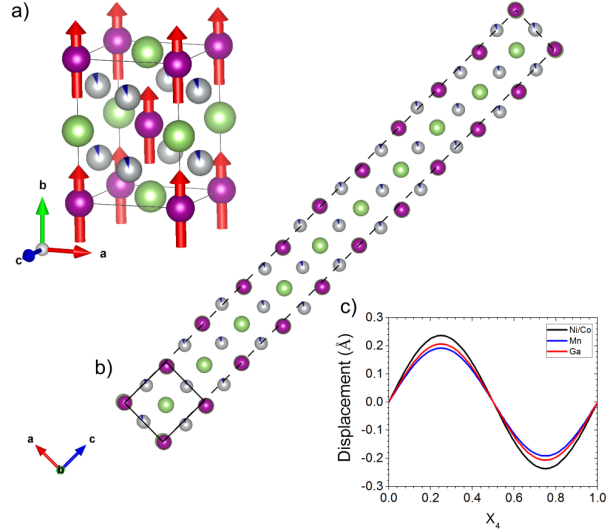

FIG. 6. a) Average nuclear and magnetic structure for the 3% Co sample in the martensitic phase at 5 K. b) Modulated structure at 5 K in the martensite structure, the solid line shows the average unit cell and the dashed lines are a guide for the eye to appreciate the sinusoidal modulation along the a direction. c) Plot of the sinusoidal displacement along x as function of the internal coordinate $x_4$.

The second transition at $T_{II}$ is characterized by a sudden jump of the cell parameters as can be seen in fig. 7. It is worth noting that the modulation vector changes its periodicity from the almost commensurate position to $q \approx (0.38\ 0.38\ 0)$ (defined in the pseudo-cubic austenite cell). The diffraction data indicate that the transition is of the first order with a clear phase coexistence as observed in the refinement of the 120 K data (see figure S1).[59] There is no evidence in the diffraction pattern of any extra reflections, ruling out the possibility of a violation of the current symmetry and the stabilization of a new modulation/superstructure. The diffraction data indicate that the transformation is driven by the tetragonal and orthorhombic lattice strains solely, excluding symmetry change and defining it as an isomorphic transition. The refinement at 5 K was then conducted in the $Im'mm'(00\gamma)s00$ superspace space group. The resulting Rietveld plot is shown in fig. 5 whereas the crystal information are available in table S VII and S IX.[59] The refined crystal structure is very similar to the martensite phase below $T_I$ with larger amplitudes of the sinusoidal modulation (twice as big) and larger strains. Very weak, second order satellites were observed, but their intensity can be reproduced with the first order harmonic modula-tion described in table S IX. [68, 69] As for the $T_I$ phase, the magnetism is mostly related to the Mn site, and the size of the moment does not change significantly with respect to the other two phases. The refinement of the moment on the Ni site converges to a finite value with statistical significance. Nevertheless, the observed moment is very small and the refinement reliability factors do not change significantly with the additional parameters. It is possible in any case to conclude that the Ni-site probably shows a ferromagnetic arrangement with respect to the Mn spins (refinement with an AFM coupling has higher reliability factor and eventually converges back to the FM alignment) with a small ordered moment around 0.2 $\mu_B$, in agreement with the electronic-structure calculation in Ref. [52].

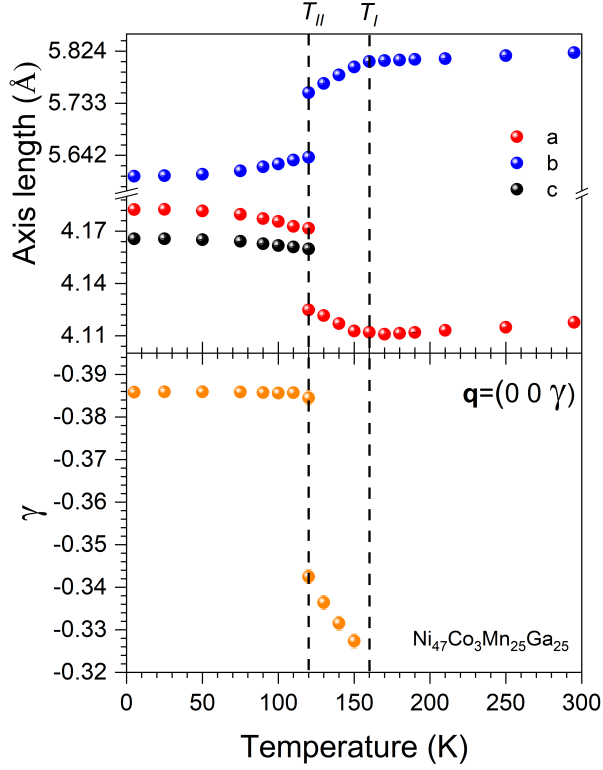

FIG. 7. Temperature evolution of the cell parameter (top) and $\gamma$ component of the modulation vector across the two martensitic transformations of the $Ni_{47}Co_3Mn_{25}Ga_{25}$ compound. The dashed lines indicate the two transition temperatures.

b. $Ni_{45}Co_5Mn_{25}Ga_{25}$  The neutron diffraction data collected on the 5% Co sample confirm the presence of a single transition at $T_{II} \approx 90$ K as observed from the magnetization data (fig. 2). Analogous to the lower doping concentration, a tetragonal splitting of the 200 reflection is evident (see fig. 8) together with the appearance of weak satellite reflections as can be seen, for instance, around the 111 reflection of the cubic austenite (see inset fig. 8). Differently form the transition at $T_{II}$ in the 3% Co sample, the change of the lattice parameter is abrupt underlining a first order transition, confirmed also by the observed phase coexistence and from the thermal hysteresis in the magnetization measurements (Fig. 2). This transition clearly resembles the $T_{II}$ transition in the 3% Co compound, but the features of the modulation are more related to the first martensitic phase observed below $T_I$ in the 3% Co sample, namely the value of the modulation vector and the weak satellite reflections. The martensite phase was refined against the data collected at 5 K in the $Im'mm'(00\gamma)s00$ magnetic superspace group. The crystal data and the atomic positions are reported in Table S X and S XI respectively,[59] whereas a Rietveld plot of the refinement at 5 K is shown in Fig. 8. As for the other martensitic phases only the sinusoidal modulation wave for each position were refined, and the obtained values are in the middle of the ones observed in the 3% Co phases. The lattice shows a slight orthorhombic distortion, and the ferromagnetic arrangement of the Mn moment is retained across the martensite transformation. The refinements yield a very small moment on the Ni/Co sublattice, but also in this case, the statistical relevance of the result is very limited. Fig. 9 shows the thermal evolution of the lattice parameters and of the modulation vector showing the first order martensitic transformation at $T_{II}$. In this case, the modulation vector at the beginning show a sharp rise followed from an almost constant value around the commensurate position 1/3.

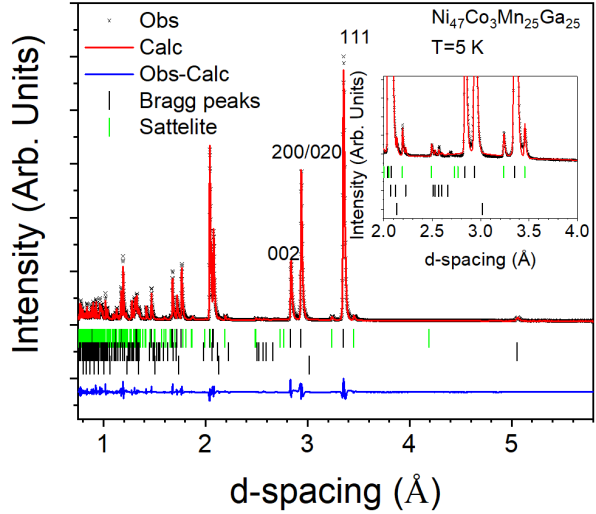

FIG. 8. Rietveld plot of the martensitic phase at 5 K in the $Im'mm'(00\gamma)s00$ magnetic superspace space group for the $Ni_{45}Co_5Mn_{25}Ga_{25}$ composition. Observed (x, black), calculated (line, red), and difference (line, blue) patterns are reported. The first tick marks row indicates the Bragg reflections position of the martensite phase with the main reflection in black and the satellite reflection in green. The second and third tick marks rows indicate the reflection from the MnO magnetic impurity and V respectively. The agreement factors are $R_p$=3.7% and $R_{wp}$=6.5%. The inset shows a zoom of the diffraction patter highlighting some satellite reflections around the 111 reflection.

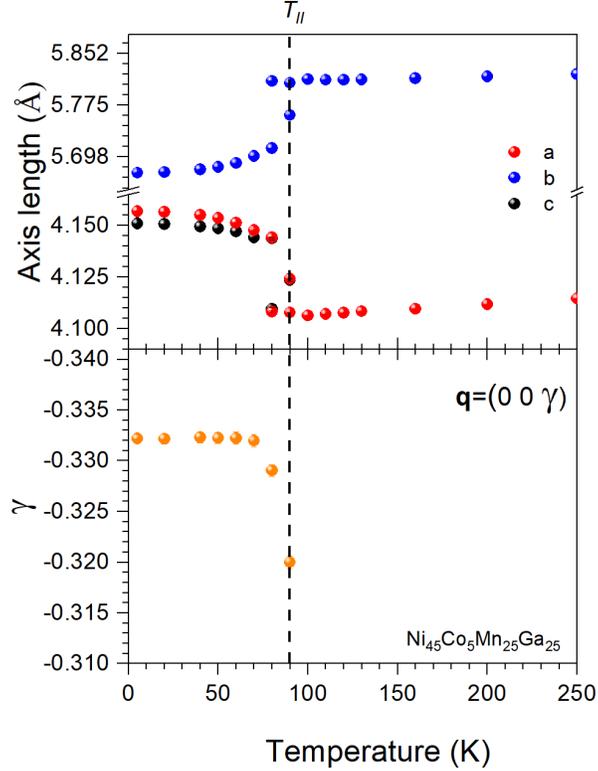

FIG. 9. Temperature evolution of the cell parameters (top) and $\gamma$ component of the modulation vector across the martensitic transformation of the $Ni_{45}Co_5Mn_{25}Ga_{25}$ compound. The dashed line indicates the $T_{II}$ transition temperature.

## IV. DISCUSSION

The high resolution neutron diffraction data hereby reported allows to gain important insights on the martensitic transformation. To obtain these information we performed a mode decomposition analysis[30, 57] of the experimentally refined structures. Such analysis supplies the decomposition of the distorted structure in terms of symmetry-adapted modes, allowing the identification of the distortions hierarchy and elucidate the driving mechanisms

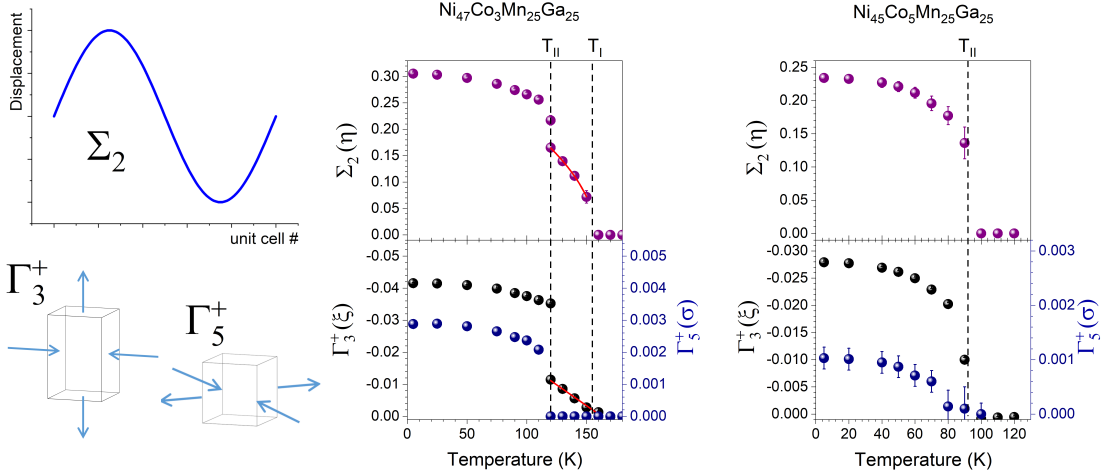

FIG. 10. Mode decomposition analysis of the $Ni_{50-x}Co_xMn_{25}Ga_{25}$ samples. In the top panels is shown the total amplitude of the $\Sigma_2$ displacive mode representative of the sinusoidal displacement of the atoms in the martensite states, the red line represents the best fit with a critical law $\propto (T - T_C)^\beta$ see text for details. The bottom panel shows the evolution across the martensitic transition of the tetragonal strain $\Gamma_3^+$ (left axis) and of the orthorhombic strain $\Gamma_5^+$ (right axis), the red line represent a linear fit of the $\Gamma_3^+$ strain close to the $T_I$ transition temperature. The dashed lines indicate the critical temperatures $T_I$ and $T_I$.

of the transitions.[30, 57]

The mode decomposition analysis,[30, 57] performed on the ferromagnetic austenite phases, indicates that the sole $m\Gamma_4^+$ mode assumes finite amplitude, which is directly proportional to the Mn and Ni magnetic moment. As pointed out previously, no changes of the magnetic structure were detected from the neutron data at the martensitic transformations indicating that the magnetic degree of freedom is not pivotal in these transitions.

The results of the mode analysis across the martensitic transitions temperatures for both compounds are shown in fig. 10. In the top panels the total amplitude of the displacive modulation $\Sigma_2$ mode is shown. This symmetry mode is related to the incommensurate sinusoidal displacement, and its amplitude is proportional to the order parameter $\eta$. On the lower panels the amplitudes of the tetragonal and orthorhombic strains are reported. These two modes are described by the $\Gamma_3^+$ and $\Gamma_5^+$ representations respectively and their amplitudes are proportional to the order parameters $\xi$ and $\sigma$. By looking to the amplitudes in fig. 10, the $\Sigma_2$ mode represents the main distortion, being six times greater than the tetragonal strain $\Gamma_3^+$ and two order of magnitude bigger than the orthorhombic strain $\Gamma_5^+$. This simple evaluation allows identifying the $\eta$ ($\Sigma_2$ mode) and the $\xi$ ($\Gamma_3^+$ modes) order parameters as the driving distortions featuring the austenite to martensite phase transition.

The temperature dependence of the two order parameters across the transitions draws an interesting scenario. In the 3% doped sample, at $T_I$, $\eta$ follows a power law dependence with a critical exponent $\beta = 0.52(5)$ (fig. 10). Interestingly, the tetragonal strain $\xi$, below $T_I$, starts to increase linearly in the same temperature-range indicating a close correlation with the modulation amplitude evolution. Specifically, the strain order parameter $\xi$ is a secondary order parameter induced by the displacive modulation $\eta$, at least in this temperature range. The two order parameters are coupled, by symmetry, with a linear quadratic invariant $\xi\eta^2$ ruling the experimentally observed temperature behavior. At $T_{II}$ in the 3% Co sample, both distortion modes present a clear discontinuity indicating a first order transition. In the 5% doped sample, a single first order transition is observed, in which the $\Sigma_2$ and $\Gamma_3^+$ modes follow the same temperature evolution, again indicating a strong coupling between the order parameters.

It is possible to rationalize these observation within the Landau phenomenological theory and show that the martensitic transformation can be described within the soft phonon model. The symmetry constrained Landau free energy potential $F(\eta, \xi)$ for the orthorhombic martensite transition is described in Appendix B and we refer to that for details. The mode decomposition clearly indicates the presence of two strongly coupled primary order parameters in the martensitic transformation. Moreover, the presence of two clear transitions in the 3% Co compound but also in the parent $Ni_2MnGa$ [25, 39] and other alloy compositions indicate that both order parameter present a clear instability, which takes place at different characteristic temperatures mainly depending on the lattice structure and chemical composition. In the 3% Co system, as well in $Ni_2MnGa$, the martensitic transition develops in two steps. First, the 'pre-martensite' phase appears: this phase transition is driven only by the displacive order parameter $\eta$, and the tetragonal strain appears as secondary order parameter as observed in the 3% Co sample as well as in Pt doped samples,[11]. The 'pre-martensite' phase in the parent compound $Ni_2MnGa$ [25, 39] do not show any measurable tetragonal strain nevertheless being allowed by symmetry it will be present. Further decrease in the temperature induces the "proper" martensite transition that is related to the strain instability.[25, 39] In both the $Ni_2MnGa$ system and in the 3% Co doped composition but also in Pt doped samples,[11] the prominent increase of the strain (see $T_{II}$ in Fig. 10) induces a renormalization of the modulation amplitude and an increase of the modulation vector component as well. It is then possible to consider both these transitions as two distinct martensitic transformations: the first one driven solely by the displacive modulation and the second one by the lattice strain.

These features can be explained by considering the linear quadratic coupling $-d_1\xi\eta^2$ present in the Landau free energy (see Appendix B) and how this effects the martensitic transition. Let's define $T_\eta$ and $T_\xi$ the characteristic temperatures of the order parameters. When $T_\xi$ is close to $T_\eta$, like in the 3 %Co compound, the $\xi\eta^2$ coupling term can re-normalizes the quadratic coefficient of the displacement order parameter as $(\frac{a_2}{2} - d_1\xi)\eta^2$ [31, 70, 71]. The $a_2$ parameter can be described within the soft phonon mode as $m^*\omega_0^2$, where omega is the frequency of the soft mode and $m^*$ is its effective mass. In the soft phonon model the frequency $\omega_0$ should continuously decrease to zero following the law $\omega_0 = [\frac{a_{2,0}(T-T_C)}{m^*}]^{\frac{1}{2}}$. Experimentally this is not the case since a first order phase transition is observed in both Co doped compounds as well as an incomplete softening of the unstable phonon is observed in the parent Ni$_2$MnGa alloys [40]. Nevertheless if we consider the presence of the linear quadratic coupling $-d_1\xi\eta^2$ and if the coupling coefficient $d_1$ is large enough the transitions become of the first order.[31, 49, 50, 70, 71] Of course there might be other reasons for the transition to be of the first order type, as explained in Appendix B, nonetheless the experimental observation of the coupling between the orders parameters indicate it as the more likely candidate to explain the incomplete softening and the first order character of the incommensurate displacement transition.

It is now interesting to analyze the situation where $T_\xi > T_\eta$. For $T > T_\xi > T_\eta$ both $a_1$ and $a_2$ parameters (see appendix B for definitions) are positive and the austenite phase is the minimum of $F(\eta, \xi)$. For $T_\eta < T < T_\xi$ the $a_1$ term became negative indicating that the thermodynamic potential has a minimum for a finite value of the tetragonal strain. In this case, if $d_1\xi < \frac{a_2(T)}{2}$ the renormalized quadratic coefficient of the displacive modulation assume a negative value and the transition to the modulated state occurs although the temperature is higher than the characteristic temperature $T_\eta$. This scenario is likely to happen in case of strongly first order transitions of the lattice strain as shown by Toledano in the case of the nuclear transition in benzyl [72] and in a more general way from Salje *et al.*[31] regarding multiferroic phase transitions. In these conditions the two strongly coupled order parameters will follow the same temperature behaviour[31] as observed in the 5% doped compound (see fig. 10 right side), but more importantly as it is been observed in the martensitic transformation of Ni-Mn-X (X=In, Sn, Sb) metamagnetic alloys.[35, 73, 74]

## V. CONCLUSION

The symmetry of the martensitic transformation in the Ni-Mn-Ga and related compounds has been discussed in detail, starting from the experimental observation of the martensitic transformation in Co doped sample of $Ni_2MnGa$ and from the description of the latter through a symmetry constrained thermodynamic potential. The analysis allows a straightforward description of the properties and characteristic of the transformation, highlighting the fundamental role played by the linear quadratic coupling between the displacement and the tetragonal strain. The experimental observation of this coupling points to a primary role of the displacive modulation in the martensitic transformation. The observation of a martensite phase in which the tetragonal strain appears as a secondary order parameter induced by the displacive modulation indicate that the latter can not be induced by the tetragonal twinning in disagreement with the adaptive model.

The $-d_1\xi\eta^2$ coupling also act as a renormalization on the displacement transition temperature explaining the strong first ordered martensitic transformation observed in the metamagnetic Ni-Mn-X composition as well as the stress induced martensite transition. It is also worth mentioning that the application of an external tetragonal strain to the austenite phase, for example as uniaxial stress along [001] cubic direction, can trigger the martensite transition to the incommensurate state thanks to the same coupling invariant, whereas the application of isotropic pressure will promote the austenitic phase. Clearly, the engineering and the control of the lattice strains become pivotal to the design of the material properties, and the stability and the composition of the austenitic lattice, in particular of the fcc sublattices, will determine the characteristic of the transformation.

### Appendix A: Magnetic symmetry of the austenite phase

By taking the $Fm\bar{3}m1'$ space group as parent structure we perform magnetic symmetry analysis, with the help of the ISODISTORT software,[56] assuming the ordering of both

TABLE I. Magnetic symmetry analysis results from the Fm-3m1' parent structure with the $m\Gamma_4^+$ irreps.

| ODP | Subgroup | Basis | Origin | Anisotropy | Cont. |
|---|---|---|---|---|---|
| P1($\mu_1$,0,0) | $I4/mm'm'$ | (0, 1/2, −1/2), (0, 1/2, 1/2), (1, 0, 0) | (0, 0, 0) | $[001]_C$ | yes |
| P2($\mu_1$,$\mu_1$,0) | $Im'm'm$ | (−1, 0, 0), (−1/2, 1/2, 0), (1/2, 1/2, 0) | (0, 0, 0) | $[110]_C$ | no |
| P3($\mu_1$,$\mu_1$, $\mu_1$) | $R\bar{3}m'$ | (1/2, 0, −1/2), (0, −1/2, 1/2), (−1, −1, −1) | (0, 0, 0) | $[111]_C$ | yes |
| C1($\mu_1$, $\mu_2$,0) | $C2'/m'$ | (0, 1, 0), (0, 0, −1), (−1/2, 1/2, 0) | (0, 0, 0) | $(101)_C$ | no |
| C2($\mu_1$, $\mu_1$, $\mu_2$) | $C2'/m'$ | (1/2, 1/2, 1), (1/2, 1/2, 0), (1/2, 1/2, 0) | (0,0,0) | $(111)_C$ | no |
| S1($\mu_1$,$\mu_2$, $\mu_3$) | $P\bar{1}$ | (1/2, 1/2, 0), (1/2, 0, 1/2), (0, −1/2, −1/2) | (0, 0, 0) | - | no |

Mn and Ni sub-lattices. The symmetry analysis leads to six possible magnetic space groups deriving from the $m\Gamma_4^+$ irreducible representation (Irreps) with different order parameter directions (ODP) (see table I). The corresponding magnetic structures are reported in fig. 11, in all the cases the relative orientation of the spins is ferromagnetic within each sub-lattice, but with different magnetic anisotropy direction and induced strain components, in agreement with the magnetic symmetry. As example in the present case the magnetic space group $I4/mm'm'$ allows the occurrence of spontaneous strains in the lattice. In particular it allows the tetragonal strain $\Gamma_3^+$ ($\xi$) that is induced from the primary magnetic mode $\mu$. This magneto-elastic coupling is described in the system free energy by the linear quadratic coupling term $\xi\mu^2$ and gives the possibility to control the spontaneous tetragonal strain with an external magnetic field as observed in magnetic-shape memory alloys.[14] In the present work this magneto-elastic coupling is not directly observed in the diffraction data but being allowed by symmetry, it will have a finite value.

**Appendix B: Landau free energy for the martensitic transformation: orthorhombic case**

The mode decomposition reported in the main text indicates two main distortions driving the two martensitic transformations. The Landau free energy for the phase transition is envisaged to be composed by these two distortions: (i) the tetragonal strain that transform as the two dimensional irreducible representation $\Gamma_3^+$ ($\xi$) (ii) the incommensurate displacive

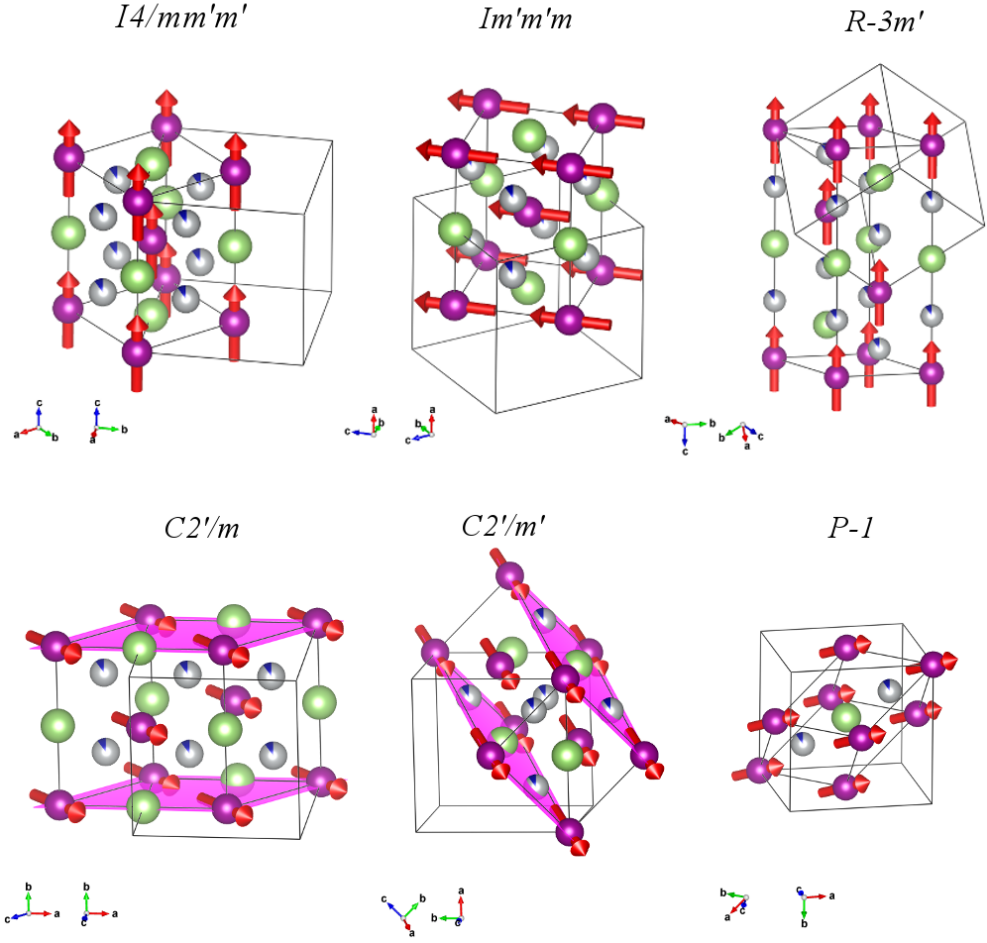

FIG. 11. Sketch of the magnetic structures derived from the parent $Fm\bar{3}m1'$ symmetry, corresponding to the action of the $m\Gamma_4^+$ Irreps having different order parameter direction and magnetic anisotropy (see table 1). The purple planes in the monoclinic magnetic space groups indicate the anisotropy

distortion that transform as the twelve dimensional $\Sigma_2$ ($\eta$) irreducible representation . In first approximation, the magnetic order parameter can be considered unchanged across the structural transition as the ordering temperature falls considerably far from the martensite transition. For these reason we exclude it from the description. The symmetry-constrained free energy can be written as follows:

$$F(\eta,\xi) = \frac{a_1}{2}\xi^2 + \frac{b_1}{3}\xi^3 + \frac{c_1}{4}\xi^4 + \frac{a_2}{2}\eta^2 + \frac{c_2}{4}\eta^4 - d_1\xi\eta^2 \tag{B1}$$

where $a_1 = a_{1,0}(T - T_\xi)$ and $a_2 = a_{2,0}(T - T_\eta)$. The $a_2$ coefficient can be related to the softening of the unstable phonon mode along the [δ δ 0] line of the Brillouin zone. Cochran [47] suggested that, close to the critical temperature, the square of the phonon frequency decrease linearly following $m\omega_0^2 \propto (T - T_C)$. Is then possible to relate the $a_2$ coefficient in equation B1 to the phonon frequency $\omega_0$ thought:

$$\omega_0 = [\frac{a_{2,0}(T - T_\eta)}{m^*}]^{\frac{1}{2}} \tag{B2}$$

In its first description the soft phonon mode described second order transition with a complete softening of the unstable phonon modes[47], but it has been shown that the model can be extended to first order transitions explaining the incomplete softening.[49, 50]. The Landau potential described in equation B1 allows the transition of the $\eta$ order parameter to be of the second order, nevertheless either the presence of coupling terms between the order parameters or $c_2 < 0$, considering the sixth degree invariant in $\eta$, can make the transition first order.

The linear quadratic term $-d_1\xi\eta^2$ defines the coupling between the displacive modulation $\Sigma_2$ and the tetragonal strain $\Gamma_3^+$. By symmetry, a second linear quadratic coupling is allowed between the displacive modulation and the orthorhombic strain $\Gamma_5^+$ ($\sigma$), but the observation of small values for the latter in most of the Heusler systems allows us to discard this term without loss of information. It is worth noting that such thermodynamic potential described here is quite common in solid-state physics,[71, 75] for example in magneto-structural transition in 3d-metal monoxides[76, 77], in the ferroelastic/ferroelectric transition in benzyl[72] or even in the crystallization of icosahedral quasicrystal.[78].

This symmetry-constrained thermodynamic potential allows three stable phases with different symmetries: the austenite parent phase (0,0) with $Fm\bar{3}m1'$ symmetry, the tetragonal non modulated martensite (0,$\xi$) described in the $I4/mmm1'$ space group corresponding to the $L1_0$ structure, and the modulated martensite phase ($\eta$,$\xi$) with the $Im'mm'(00\gamma)s00$ symmetry.

**ACKNOWLEDGMENTS**

The authors acknowledge the Science and Technology Facilities Council for providing neutron beamtime on the WISH beamline. FO acknowledge Dr. Giovanni Romanelli for useful discussion.This work has benefited from the COMP-HUB Initiative, funded by the

---